\documentclass{iauc}
\usepackage{graphicx}

\title[Impact of accurate distances to dEs on cosmology] 
{The impact of accurate distances to dwarf elliptical galaxies on cosmology}
\author[Helmut Jerjen]   
{Helmut Jerjen}

\affiliation{Research School of Astronomy and Astrophysics, Australian National University,
Australia \break 
email: jerjen@mso.anu.edu.au}

\pubyear{2005}
\volume{198} 
\pagerange{}
\date{}
\jname{Near-Field Cosmology with Dwarf Elliptical Galaxies}
\editors{H. Jerjen and B. Binggeli, eds.}
\begin{document}

\maketitle

\begin{abstract}
The Surface Brightness Fluctuation method has been shown to be a powerful distance 
indicator for dwarf elliptical galaxies to very low surface brightness levels. It is applicable 
to stellar systems that are out of reach for
classical indicators requiring resolved stellar populations such as the tip magnitude 
of the red giant branch. I briefly discuss a few results from recent SBF studies 
of dEs  to demonstrate the significance of the SBF method to address 
long-standing issues related to cosmography, dark matter in galaxy groups, substructures 
in clusters, and the discrepancy between the mass function of collapsed objects and the faint 
end of the galaxy luminosity function. For the analysis of the large number of galaxy 
images that need to be processed as part of such SBF studies we are currently developing a fast, 
semi-automatic reduction pipeline that will be made readily available to the astronomical community.
\end{abstract}

\firstsection % if your document starts with a section,
\section{Motivation}
The ability to spatial locate newly detected dwarf galaxy candidates is of fundamental
importance as it enables us among others to allocate them to an environment and to 
measure physical quantities such as their luminosity, mass, and star formation history. For instance 
any dwarf galaxy that can be found either in or close to the Local Group is of  greatest interest. Such 
stellar systems are relatively rare and due to their proximity represent prime targets 
for detailed stellar population and dark matter studies. That explains why the scientific interest 
in Andromeda IV,  discovered during van den Bergh's (1972) search for dwarf spheroidal 
companions to M31, was initially high but depreciated quickly  when a radial velocity and a tentative 
distance measurement revealed 28 years later (Ferguson, Gallagher \& Wyse 2000) that it 
actually is a background irregular galaxy well outside the confines of the Local Group. 
 
But before a distance can be measured the dwarf galaxy needs to be detected first. A modern and probably the most promising way to find gas-rich {\it dwarf irregular} (dIrr) galaxies in the local Universe 
is via deep, blind HI surveys using newly developed multi-beam instruments at radio 
telescopes. Because low luminous and low surface brightness dIrrs have increasingly large  HI\,mass 
fractions (e.g.~Warren et al.~2004), simultaneous detection and 
distance measurement via redshift remains possible at 21\,cm even in extreme cases like 
DDO154 (Krumm \& Burstein 1984; Carignan \& Freeman 1988) or ESO215-G?009 
(Warren et al.~2004) where only an elusive stellar component is present, embedded in 
an extended disk of neutral hydrogen that has not been involved in the galaxy's 
star-formation processes. 

Evidence for a high detection completeness with this search method is provided by the HI Parkes 
All-Sky Survey (HIPASS, Barnes et al.~2001). Perhaps somewhat surprising, HIPASS 
found {\it no genuine dark galaxies, ie. extragalactic HI clouds or dark matter halos filled 
with HI but no stars}  above the detection limit of  $M_{HI}>10^7\,M_\odot$. All  but a very 
small number of the $\sim$6000 HIPASS detections in the surveyed velocity range 
$-700$\,km\,s$^{-1}<v_\odot <12,700$\,km\,s$^{-1}$ could be associated with stars 
(Koribalski et al.~2004, Doyle et al. 2005). The few exceptions are tidal features 
(e.g.~Ryder et al.~2001, Oosterloo et al.~2004) or high velocity clouds (Kilborn et al.~2000). 

Finding {\em dwarf elliptical} (dE) galaxies is by nature a more difficult task. 
Karachentsev with collaborators and many others  identified large numbers of 
dE candidates in the direction of nearby groups and clusters over the last decade using primarily 
wide-field photographic plates. The challenge remains to measure distances to all 
of them to separate the wheat from the chaff. Unlike in the case of the 
dIrrs, the absence of gas in dEs prevents a relative simple distance estimate via redshift. 
Moreover, the large scatter in well-known empirical scaling relations based on 
integral galaxy parameters such as the surface brightness--luminosity (Binggeli \& 
Cameron 1991) or the shape parameter--luminosity relation 
(Binggeli \& Jerjen 1998) makes them unsuitable for individual distance measurements.  

That is why in the last decade, a formidable observational effort has been devoted to derive distances 
to dEs by using the tip magnitude of the red giant branch (TRGB), a powerful Pop II distance indicator 
(Karachentsev et al.~2004 and references therein). But the requirement of accurate photometry 
for individual stars 1--2 magnitudes below the tip sets a practical application 
limit at $\approx$ 5\,Mpc. TRGB measurements for galaxies beyond that limit become 
quickly very costly and time consuming as was demonstrated by 
Harris et al.~(1998) who spent  9 hours with the HST to obtain the TRGB distance for 
one single dE (IC3388) in  the Virgo cluster. 

Because of this limitation, a relatively new distance indicator the so-called surface brightness 
fluctuation (SBF) method has received increasing attention in recent years. The method 
quantifies the statistical pixel-to-pixel variation of star counts across a galaxy image with 
the major technical advantage of working on unresolved stellar populations.
Ground-based CCD images with relative short integration times can be used instead of 
high resolution space-based observations which makes the SBF method extremely efficient. 
Distance measurements for early-type galaxies far beyond the reach of any other distance 
indicator become possible. 

The theoretical framework of the SBF method was developed by Tonry \& Schneider (1988) 
with one of the main aims to study the distribution and kinematical properties of large-scale structure 
outlined by luminous giant ellipticals in the local Universe (e.g.~Tonry et al.~2001). 
While SBF applications were most exclusively focussed on these high surface brightness 
systems, Jerjen and collaborators (Jerjen, Freeman \& Binggeli 1998, 2000; Jerjen et al.~2001, 
Rekola, Jerjen \& Flynn 2005) 
demonstrated that the method works equally well for low surface brightness dEs as faint 
as $\mu_{\rm B, eff} =26$\,mag\,arcsec$^{-2}$ and $M_B= -10$\,mag. 
With this possibility to measure distances to large numbers of dEs easily, including 
the newly identified dE candidates, a special opportunity arises to make progress in areas most 
relevant to near-field cosmology including
\begin{itemize}
\item Mapping the gravitational centers in the Supergalactic plane and Local supercluster. 
\item Substructures and dark matter contents of galaxy clusters. 
\item dE density in the local Universe.
\item Shape of the galaxy luminosity function to the lowest luminosities. 
\end{itemize}

\section{SBF calibration}
As the strength of the SBF signal and the derived fluctuation magnitude $\overline{m}$ 
depend not only on the distance of the galaxy but also on its stellar composition, the 
calibration of the method must take into account the basic properties of the stars. For a 
first theoretical calibration of the SBF method in the $R$-band, Jerjen et al.~(1998, 2000) 
studied the effect of different mixtures of ages and metallicity as well as the starformation 
history on the  observable quantities, fluctuation magnitude $\overline{M}_R$ and $B-R$ colour.
For that purpose, Worthey's (1994) models combined with Padova isochrones 
(Bertelli et al.~1994) were employed to simulate a range of synthesis stellar 
populations with two starformation bursts to make allowance  for the variety of
starformation histories observed among Local Group dEs/dSphs (Grebel 2001).
The age of the first burst was set at 8, 12, or 17 Gyr with a metallicity 
in the range from $-2.0$ to $0$. We added a second population, 5 Gyr 
old and of solar metallicity, that contributed 0, 10, 20, or 30\% in mass to the entire 
population. In all cases we assumed a Salpeter IMF. The ($B-R, \overline{M}_R$) values 
for a total of 170 synthetic stellar populations were calculated.

\begin{figure}[t]
\begin{center}
\includegraphics[height=8cm]{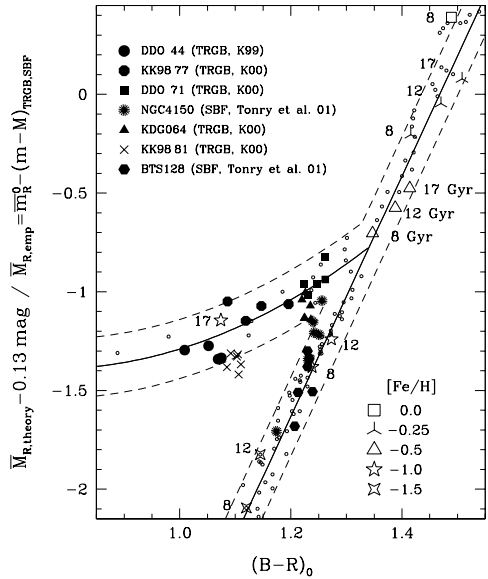}
\caption{The SBF calibration diagram for the $R$-band. The open symbols are the 
($B-R, \overline{M}_R$) values for 170 synthetic stellar populations based on Worthey (1994) 
models and Padova isochrones (Bertelli et al.~1994). An offset correction of 0.13\,mag is 
applied to the theoretical values to match the empirical results from 36 SBF fields 
measured  in 7 nearby dwarf galaxies with independent TRGB and SBF 
distances (filled symbols). The solid lines represent the best analytical functions describing 
the loci of the models while the dashed lines indicate the $\pm 0.15$\,mag uncertainty 
bands.}\label{fig:calibplot}
\end{center}
\end{figure}

The theoretical predictions were then tested with 36 SBF measurements in 
seven nearby galaxies (Jerjen et al.~2001, Rekola, Jerjen \& Flynn 2005) for which independent 
distances were available (Karachentsev et al.~1999, 2000; Tonry et al.~2001). Good 
agreement was found after a systematic offset correction of 0.13\,mag was applied to the theoretical
values.These corrected ($B-R, \overline{M}_R$) data points are shown in Fig.~1, indicated 
by the various open symbols. The two lines are best analytical fits of the empirically anchored  theoretical data and represent the calibration equations for the SBF method in the $R$-band currently available (Jerjen et al.~2001). The typical overall uncertainty in a SBF distance 
measurement is $\approx 10$\%  which includes the intrinsic scatter of the Worthey+Padova 
model values and an uncertainty of 0.1\,mag in the TRGB zero point.

 Calibrations for other filter sets have been established by various authors  
 (e.g.~$V$, $I$ filters, Tonry et al. 1997, Mieske \& Hilker 2003; SDSS $g$, 
 $z$ filters, Mei et al.~2005). 
 
\section{Some SBF applications of cosmological relevance}
\subsection{3D-mapping of cluster cores}
There is a growing interest in the structural analysis of galaxy clusters 
to evaluate the importance of external mechanisms driving galaxy evolution 
in different kinds of environments, from cluster cores to their outskirts. In this 
context the characterization of the physical conditions at the immediate location 
of a cluster galaxy (distance) and the galaxy's dynamical status (velocity)
are essential quantities.

Methods for detecting cluster substructure classically resort to projected 
galaxy positions and redshifts. But as redshifts are generally available only 
for high surface brightness galaxies, Es, S0s, and spirals where the latter 
are no genuine  cluster galaxies and the number of Es+S0s in a 
cluster is relatively small, samples tend to be dominated by late-type galaxies 
that are infalling at a late stage of the cluster evolution and thus are kinematically 
distinct from genuine cluster galaxies. Consequently, these studies  preferencially  
probe substructures in the cluster outskirts which are due to directional,
inhomogenous galaxy aggretion along the larger-scale filaments a cluster is embedded in.

The detection of substructures/group remnants in the densest cluster regions where early-type
galaxies dominate is more difficult as this is hampered by the small number of available giant 
ellipticals 
and the general lack of redshift information for dEs,  the largest galaxy population that resides 
long enough in the cluster to trace the gravitational potential. Accurate distances of a 
sufficient number of early-type galaxies are indispensable for such studies. 

In a pilot project, Jerjen, 
Binggeli \& Barazza (2004) measured SBF distances to 16 luminous dEs and 
combined them with SBF distances to giant ellipticals from the literature (Neilsen 
\& Tsevtanov 2000, Tonry et al.~2001) to trace the gravitational centers of the 
Virgo cluster along the line-of sight. The sample dEs showed a substantial spread in 
distances ranging from 14.9 to 21.3\,Mpc (see Fig.~2) which confirmed that the 
Virgo cluster core has an elongated shape, consistent with the previous findings of Neilsen 
\& Tsvetanov (2000), West \& Blakeslee (2000) and Arnaboldi et al. (2000). 
However, somewhat unexpected was that this seemingly dynamically relaxed system of 
early-type galaxies were clumped in two major concentrations at $(m-M)=31.02$\,mag 
(15.8\,Mpc) and  at 31.33\,mag (18.5\,Mpc) with a number ratio of 3:1 (Fig.~2). 
An adaptive kernel analysis of the distribution found that these clumps are 
significance at the 99\% and 89\% level, respectively, and that there might be 
even a minor third clump of galaxies at $\approx 31.55$\,mag.

\begin{figure}[t]
\begin{center}
\includegraphics[height=7cm]{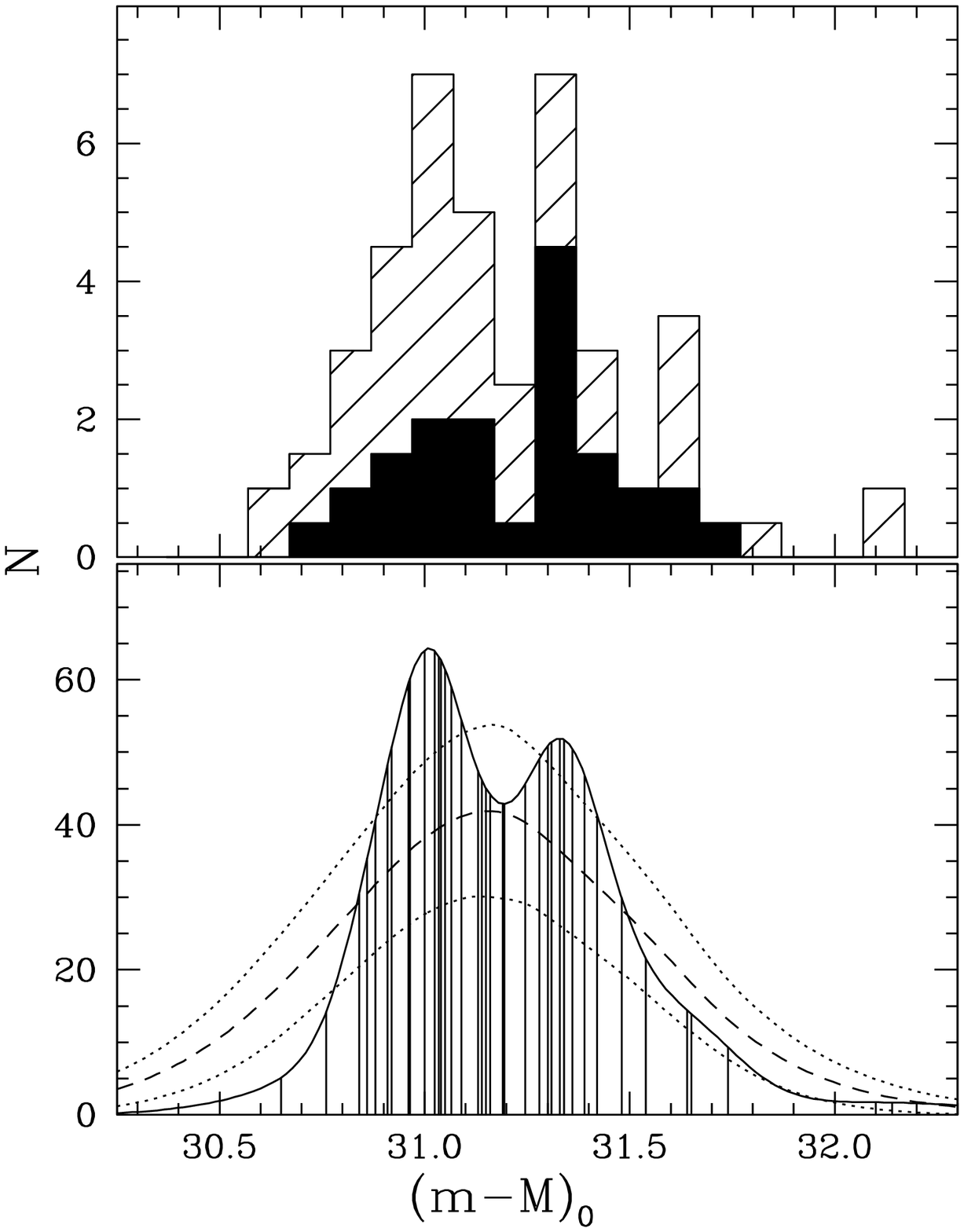}
\includegraphics[height=7cm]{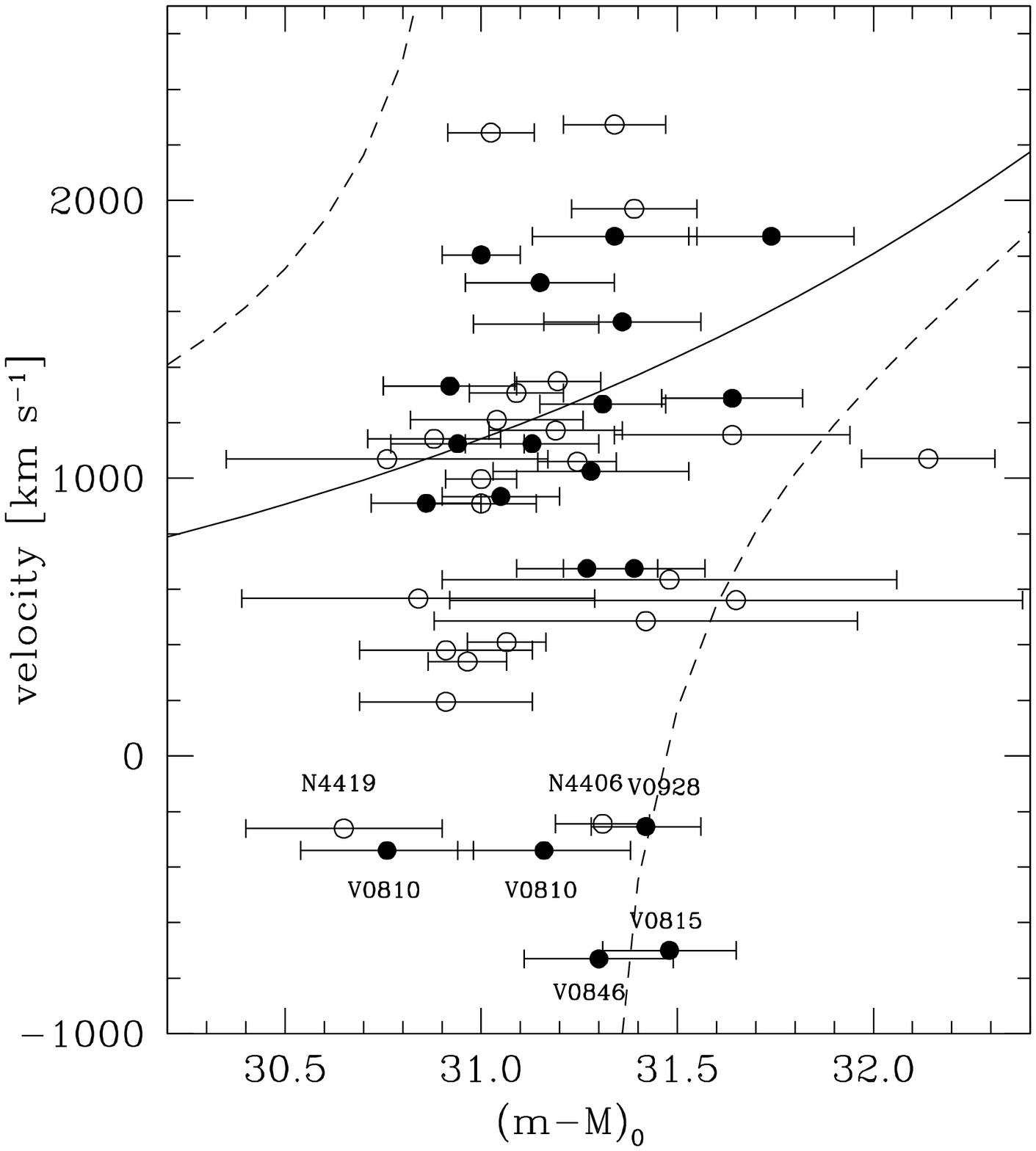}
 \caption{(Left-top): distance distribution of 41 early-type dwarf (solid histogram) and 
giant (hatched histogram) galaxies in the Virgo cluster. (Left-bottom): the same distribution
based on the adaptive kernel method allows the quantification of the apparent bimodal structure.
(Right): the distance-velocity plot (black: dEs, circle: Es) identifies the four highly blueshifted Virgo galaxies VCC810, 815, 846, and 928 as a separate group centered around M86 infalling from the 
backside of the cluster. VCC810 is shown twice to indicate its two possible SBF distances (For more details see Jerjen, Binggeli \& Barazza 2004).}
\end{center}
\end{figure}

By combining the newly obtained SBF distances with redshift information 
from the literature the long-standing mystery around the four highly blueshifted 
dE galaxies VCC810, 815, 846, and 928 could be solved. The distance-velocity 
map (Fig.~2) revealed that these systems are part of the second clump 
approximately $2.7$\,Mpc at the back of the major galaxy concentration at 15.8\,Mpc. 
Together with M86, these dEs constitute a dynamically independent group  
infalling into the cluster from the backside for the first time. 
These dwarf systems might have been late-type disk galaxies 
which, by gravitational interactions with the ICM and other cluster members, 
lost their gas and transformed into the spheroids we see today (Mao \& Mo 1998, 
Moore et al.~1998, Mastropietro et al. 2005). 
Therefore, they present prime targets for follow-up studies to search for HI cloud 
remnants and structural features as have been already found in other Virgo
dEs (Jerjen, Kalnajs \& Binggeli 2000; Barazza, Binggeli \& Jerjen 2002) and  
to test environmentally induced transformation processes such as the 
harassment scenario developed by Moore et al.~(1998).

More results on the spatial distribution of Virgo Es and dEs  can be 
expected from the ACS-Virgo Survey (C\^ot\'e et al., this volume) where SBF 
distances to $\approx 100$ early-type galaxies will be reported. The ACS Survey
is still a targeted survey. To handle the even larger volumes of CCD images of 
cluster galaxies produced by ground-based wide-field camera surveys 
at telescopes like {\it Subaru, VISTA, CFHT}, and {\it VST} in the 
near future, we are currently developing a fast SBF reduction package 
(Dunn \& Jerjen, this volume).

\subsection{Towards a distance-based galaxy luminosity function}
It is becoming increasingly evident that there are fewer dwarf galaxies per
giant galaxy observed in the local Universe than anticipated by the Cold Dark 
Matter (CDM) hierarchical clustering models (Klypin et al. 1999; Moore et al. 1999; 
Trentham \& Tully 2002). This problem may well have a complex solution: variations 
on CDM cosmology, such as Warm Dark Matter (Bode et al.~2001) or Self-Interacting 
Dark Matter (Spergel \& Steinhardt 2000). Alternatively, there is no shortage of 
astrophysical mechanisms that can diminish the accumulation of baryons in low 
mass potential wells: long cooling times for primordial gas in small halos (Haiman 
et al.~1996), galactic winds driven by supernovae and hot stars (Dekel \& Silk 1986), or
pressure support against collapse of the intergalactic plasma after reionization 
(Thoul \& Weinberg 1996; Gnedin 2000).  

The present situation is confused because there are a multitude of
reasons why the faint end of the galaxy luminosity function that is completely governed
by the {\it dwarf} galaxies (Binggeli, Sandage \& Tammann 1988) 
could deviate from the simple CDM theory expectation. Progress on this important 
problem is currently limited by observations, not theory. Most recent studies of the galaxy 
luminosity function in Virgo (Phillipps et al.~1998,Trentham \& Hodgkin 2002) and 
Fornax (Kambas et al.~2000, Hilker  et al.~2003) showed that {\it the ambiguity in 
attributing membership status to cluster/group galaxy candiates is the ultimate
source of uncertainty} on the quest to find the accurate shape, slope, and possible 
turning point of the faint end of the galaxy luminosity function. To achieve an undisputable
separation between members of an environment and background/foreground objects 
distances are needed.

\begin{figure}
\begin{center}
\includegraphics[height=3.8cm]{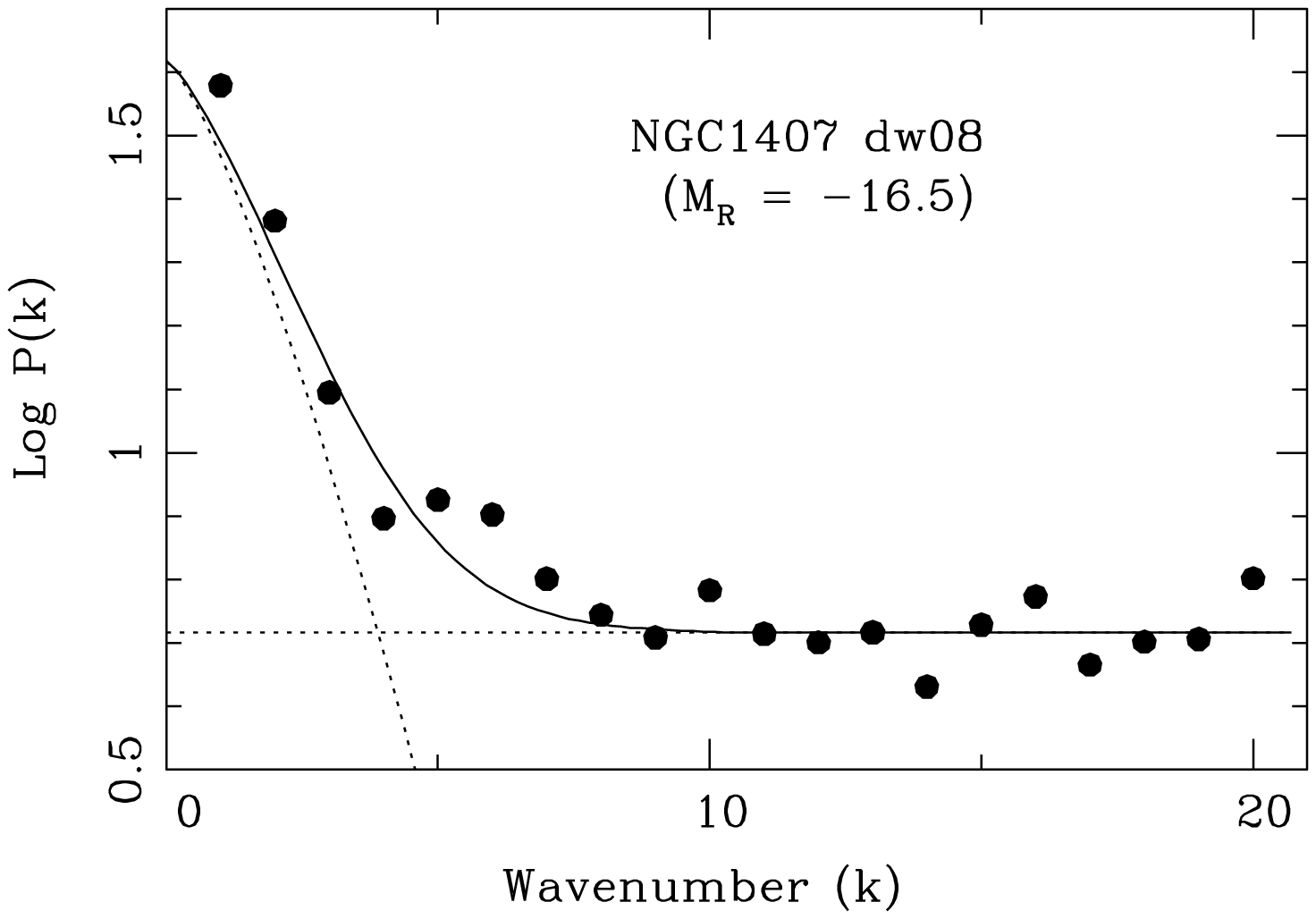}
\includegraphics[height=3.8cm]{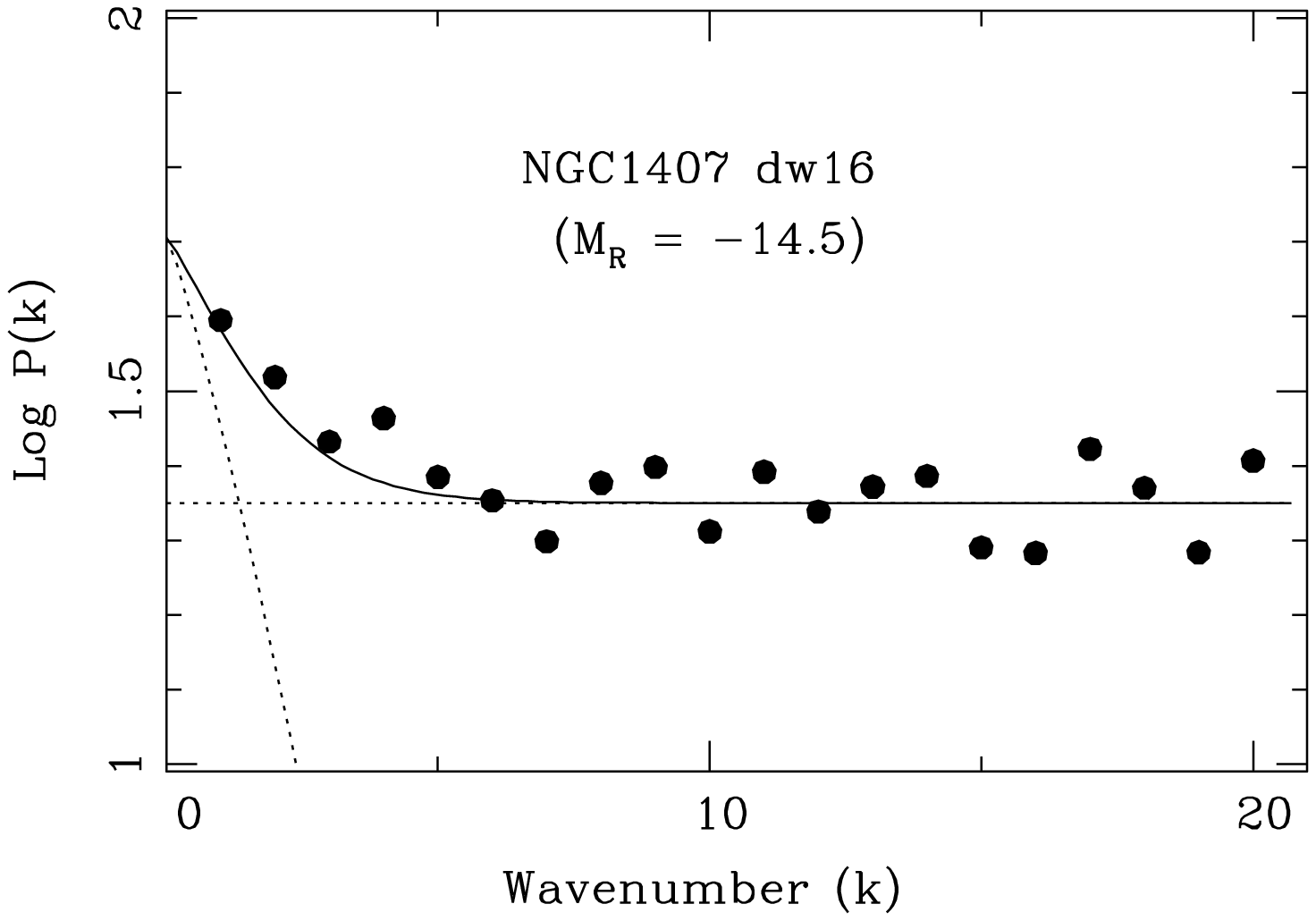}

\includegraphics[height=3.8cm]{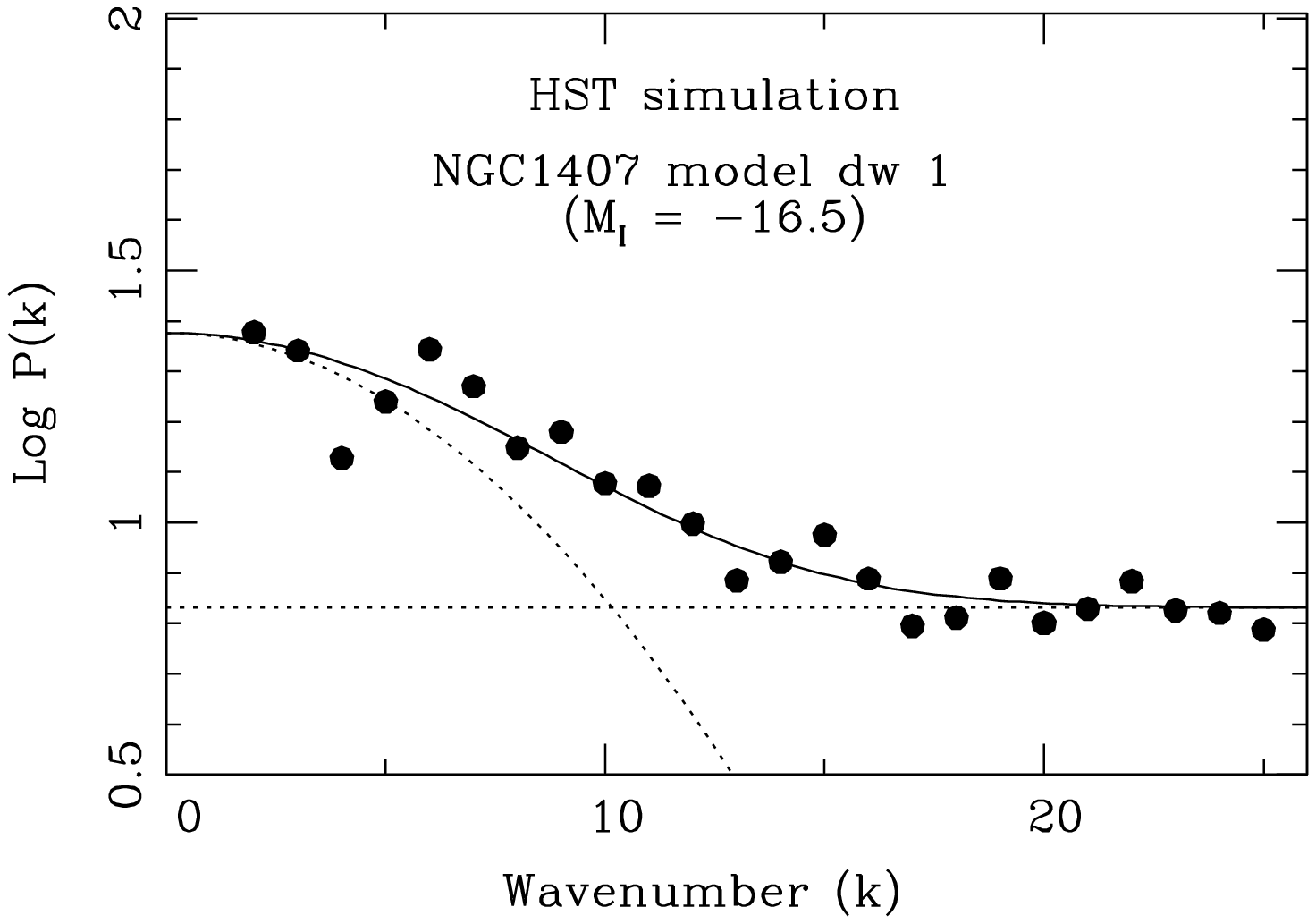}
\includegraphics[height=3.8cm]{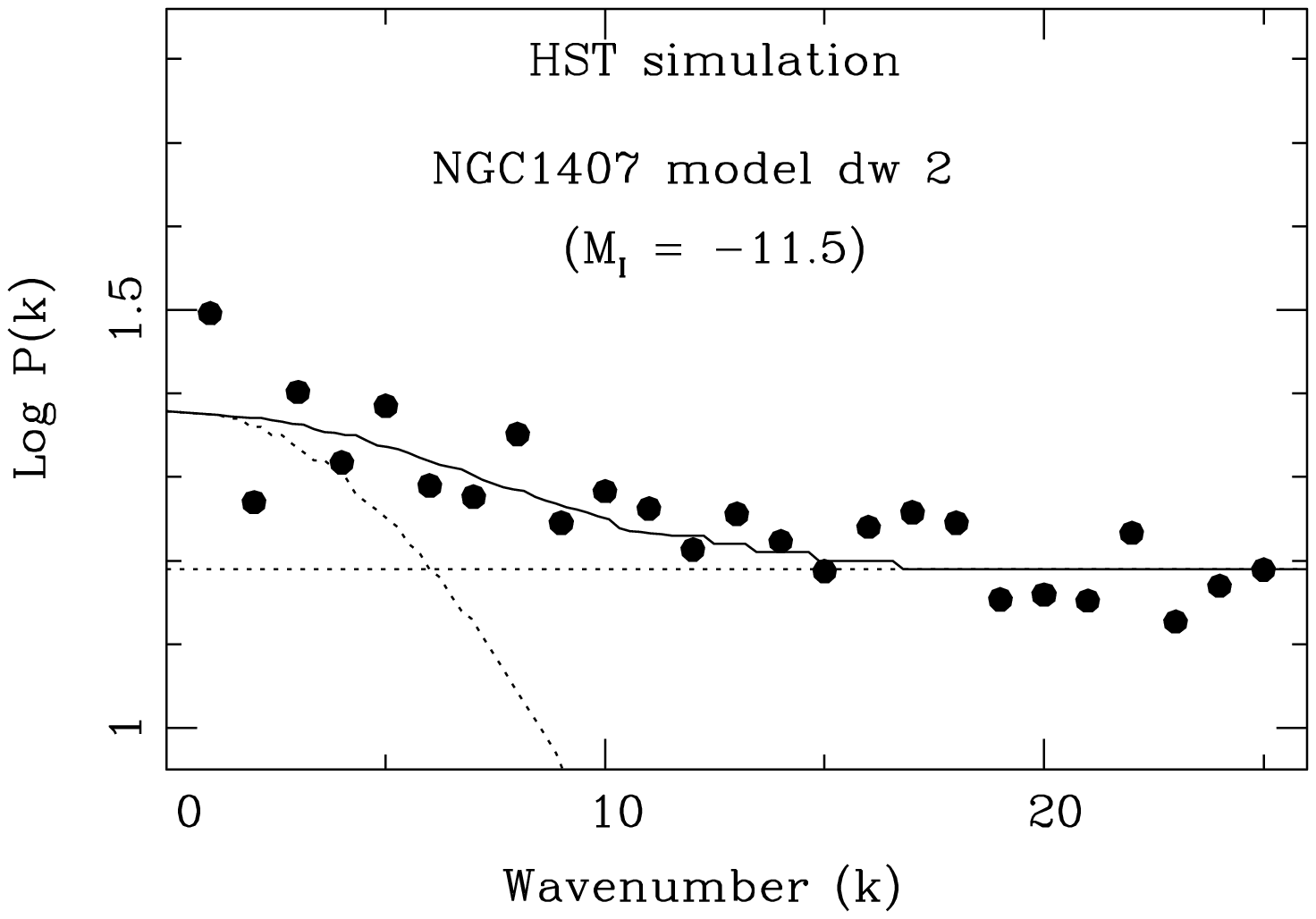}
\caption{The top panels show the $R$-band power spectra of two dwarf galaxies 
at the  the NGC1407 distance of 25\,Mpc obtained with 12 minute exposures in 
$1"$ seeing with the 8m Subaru Telescope.  The bottom panels show simulated 
$I$-band power spectra of NGC1407 dwarfs as would be obtained with 
20 minute exposures with HST/ACS.  
The absolute magnitudes of the targets are indicated in the panels.  The brighter and fainter 
simulated targets have central surface brightnesses of 22.5 and 25\,I\,mag\,arcsec$^{-2}$ 
respectively. The observations and simulations (filled circles) are well fitted by the sum 
(solid line) of a scaled version of the power spectrum of the point spread function and a 
constant (dashed lines). We note that $R-I$ is generally small ($\approx 0.5$) for early-type 
dwarfs and thus the results from the two filters comparable.}
\end{center}
\end{figure} 

That is why we are currently exploring the possibility to verify membership of all known 
232 group candidates of the NGC1407 Group (Trentham \& Tully 2002) by means of 
measuring SBF distances using high quality images from Subaru, CFHT, and HST. 
This galaxy group at $\approx 25$\,Mpc  is quite noteworthy. 
Dense knots of early-type galaxies are found across a considerable scale 
range from rich clusters with in excess of $10^{15} M_{\odot}$ to just a few 
big galaxies in a halo of $10^{13} M_{\odot}$ (Tully 2005).  Within the domain 
$V < 3000$~km\,s$^{-1}$, the NGC 1407 Group is the best case at the low mass end 
of this range of E/S0 knots.  It contains only two $L^{\star}$ galaxies, yet the 1-D velocity
dispersion is with 385~km\,s$^{-1}$ comparable to the Fornax cluster.  It contains
many dE dwarfs.  The goal of our extensive SBF study is to obtain a complete inventory.

Subaru and CFHT images can provide SBF signals satisfactory for 
distance estimates for the 63 NGC1407 Group candidates that are brighter 
than $M_R \sim -14$\,mag. Examples of the power spectra obtained for two 
galaxies are shown in the top panels of Fig.~3. The remaining 169 low luminous, 
low-surface brightness dwarfs previously unaccessible to distance and redshift
measurements, will be target objects in our HST SNAP program. 
Simulated power spectra for dwarfs as faint as $M_I = -11.5$\,mag are shown in the 
lower panels of Fig.~3.  The fluctuation signal remains above the
noise level out to wave number $k \sim 15$.  Indeed, the SBF detection 
with HST/ACS of a dwarf with $M_I = -11.5$\,mag will be better than the SBF 
detection with the current Subaru data of the galaxy with $M_R = -16.5$\,mag.  
Based on these simulation predictions we will comfortably be able to 
measure SBF signals and distances for our faintest candidates.

The results from this SBF study will enable us to establish a 
distance-based galaxy luminosity function for that particular group environment down to 
$M_R = -11$\,mag. Moreover, the feedback from the NGC1407 Group will 
then be combined with spectroscopic and neutral hydrogen data measured for the 
few late-type group member candidates to allocate accurate membership probabilities to 
the morphology-based five-staged rating scheme employed by Trentham \& Tully (2002). 
Transferring the calibrated rating scheme to other galaxy groups under study will then
help to much better constrain the luminosity functions found in other environments, permitting 
progress on the important question of whether there are differences with environment.

\section{Conclusion}
With the brief discussion of some recent projects and results based on 
SBF distances of early-type dwarf galaxies I hope I was able to let readers 
discover the great potential of this powerful dE distance indicator. Due to its simplicity 
and efficiency there is little doubt that this method will play an important role in the future
3-D surveying of the local universe, particularly taking advantage of the deep, high quality 
CCD imaging data  produced by the steadily increasing number 
of wide-field cameras on present and new generation telescopes.

\begin{acknowledgments}
HJ would like to express his gratitude to the IAU for the financial support of this most 
fruitful and enjoyable meeting. Part of my research on dwarf galaxies is supported by the 
ARC Discovery Grant DP0343156. 
\end{acknowledgments}

\begin{discussion}

\discuss{Nelson Caldwell} {It's interesting that the dwarfs you have observed so far
fall either on the linear or parabolic part of the $\overline{m}$ vs.~B-R relation.
Naively, one might expect a range of star formation histories in galaxies
and so a scattering of points between the two curves. With more objects 
observed, do you think the dEs will fill in the area between the curves?}

\discuss{Helmut Jerjen}{A possible explanation for this empty region is 
that the colour of the RGB is largely independent of age but is strongly 
dependent on metallicity (Fig.~1 of Da Costa 1997) and that for a 
fixed metallicity the colour of the HB has a discontinuity from blue for the 
oldest populations ($\sim$15\,Gyr) to red at younger ages ($\leq$10\,Gyr; 
Fig.~4 of Da Costa 1997). The latter effect is more marked for lower metallicities. These 
two trends describe the behavior of $(B-R)$ and $\overline{M}_R$ in Fig.~1. In the 
models, only the very old and metal-poor populations are located on the 
parabolic curve. With an increasing fraction of younger stars the total population
gets redder and the model values reach quickly the linear curve.}

\discuss{Steffen Mieske}{How do you know that the models you are using for the SBF 
calibration are correct?}

\discuss{Helmut Jerjen}{In Jerjen, Freeman \& Binggeli (1998) we compared the 
Worthey+Yale and Worthey+Padua results with empirical data and found much better 
agreements with the latter models. This good agreement still exists after using the 
latest SBF data for seven calibration galaxies shown in Fig.~1. But ultimately we are aiming  
for a model-independent SBF calibration based solely on nearby dEs.  
}

\discuss{Igor Chilingarian} {For the color-metallicity relation you assume old ages 
($> 8$\,Gyr), but in the paper of Geha et al. who observed several tens of dEs in
Virgo the  mean age is estimated 3 Gyr (using Lick indices). How will this 
affect the relation?}

\discuss{Helmut Jerjen}{
The $(B-R, \overline{M})$ data points for stellar populations of younger ages ($<8$\,Gyr)
would move down the linear branch in Fig.~1 blending with the results for 
older stellar populations but lower metallicities.
Caldwell, Rose \& Concannon (2003) also tried  to disentangle the effects of age 
and metallicity in a sample of Virgo early-type galaxies. The Balmer lines
in these dwarfs were interpreted within the Lick index system as being 
primarily caused by young age, rather than by low metallicity leading to
galaxy ages ranged from 1--16 Gyr. This interpretation stands in contrast 
to Maraston \& Thomas (2000, 2001), who showed that a mix of an old
metal-rich and an old metal-poor component can produce strong Balmer 
and metal lines without invoking a young population.
I also would like to recall another result from
age-sensitive narrow-band photometry in the modified
Str\"omgren filter system by Rakos et al. (2001). These authors
derived a mean age of $10\pm1$ Gyr for a sample of 27 Fornax
dEs.
}

\end{discussion}
\end{document}